\documentclass[amsmath,amssymb,amsfont,aps,prl,superscriptaddress,twocolumn, notitlepage, nobibnotes,nofootinbib,longbib,showkeys, 10pt]{revtex4-2}

\usepackage{physics}
\usepackage{graphicx}
\usepackage{amsmath,bm}
\usepackage{booktabs}
\usepackage{url}
\usepackage{soul}
\usepackage{array}
\usepackage[colorlinks,citecolor=blue,linkcolor=blue,urlcolor=blue, breaklinks=true]{hyperref}

\newcommand{\comment}[1]{{#1}}

\begin{document}
\title{Experimental quantum-enhanced kernels on a photonic processor}

\author{Zhenghao Yin}
\email{zhenghao.yin@univie.ac.at}
\affiliation{University of Vienna, Faculty of Physics, Vienna Center for Quantum
Science and Technology (VCQ), Boltzmanngasse 5, Vienna A-1090, Austria}
\affiliation{University of Vienna, Faculty of Physics, Vienna Doctoral School of Physics (VDSP), Boltzmanngasse 5, Vienna A-1090, Austria}

\author{Iris Agresti}
\email{iris.agresti@univie.ac.at}
\affiliation{University of Vienna, Faculty of Physics, Vienna Center for Quantum
Science and Technology (VCQ), Boltzmanngasse 5, Vienna A-1090, Austria}

\author{Giovanni de Felice}
\affiliation{Quantinuum, 17 Beaumont Street, Oxford OX1 2NA, UK}

\author{Douglas Brown}
\affiliation{Quantinuum, 17 Beaumont Street, Oxford OX1 2NA, UK}

\author{Alexis Toumi}
\affiliation{Quantinuum, 17 Beaumont Street, Oxford OX1 2NA, UK}

\author{Ciro Pentangelo}
\affiliation{Dipartimento di Fisica, Politecnico di Milano, piazza L. Da Vinci 32, 20133 Milano, Italy}
\affiliation{Istituto di Fotonica e Nanotecnologie, Consiglio Nazionale delle Ricerche (IFN-CNR), piazza L. Da Vinci 32, 20133 Milano, Italy}

\author{Simone Piacentini}
\affiliation{Istituto di Fotonica e Nanotecnologie, Consiglio Nazionale delle Ricerche (IFN-CNR), piazza L. Da Vinci 32, 20133 Milano, Italy}

\author{Andrea Crespi}
\affiliation{Dipartimento di Fisica, Politecnico di Milano, piazza L. Da Vinci 32, 20133 Milano, Italy}
\affiliation{Istituto di Fotonica e Nanotecnologie, Consiglio Nazionale delle Ricerche (IFN-CNR), piazza L. Da Vinci 32, 20133 Milano, Italy}

\author{Francesco Ceccarelli}
\affiliation{Istituto di Fotonica e Nanotecnologie, Consiglio Nazionale delle Ricerche (IFN-CNR), piazza L. Da Vinci 32, 20133 Milano, Italy}

\author{Roberto Osellame}
\affiliation{Istituto di Fotonica e Nanotecnologie, Consiglio Nazionale delle Ricerche (IFN-CNR), piazza L. Da Vinci 32, 20133 Milano, Italy}

\author{Bob Coecke}
\affiliation{Quantinuum, 17 Beaumont Street, Oxford OX1 2NA, UK}

\author{Philip Walther}
\email{philip.walther@univie.ac.at}
\affiliation{University of Vienna, Faculty of Physics, Vienna Center for Quantum
Science and Technology (VCQ), Boltzmanngasse 5, Vienna A-1090, Austria}
\affiliation{Christian Doppler Laboratory for Photonic Quantum Computer, Faculty of Physics, University of Vienna, 1090 Vienna, Austria}

\begin{abstract}

Recently, machine learning had a remarkable impact, from scientific to everyday-life applications. However, complex tasks often imply unfeasible energy and computational power consumption. 
Quantum computation might lower such requirements, although it is unclear whether enhancements are reachable by current technologies. 
Here, we demonstrate a kernel method on a photonic integrated processor to perform a binary classification.
We show that our protocol outperforms state-of-the-art kernel methods including gaussian and neural tangent kernels, exploiting quantum interference, and brings a smaller improvement also by single photon coherence.
Our scheme does not require entangling gates and can modify the system dimension through additional modes and injected photons. 
This result opens to more efficient algorithms and to formulating tasks where quantum effects improve standard methods.

\end{abstract}


\maketitle


\section*{Introduction}

The past decades have witnessed a swift development of technologies based on quantum mechanical phenomena, which have opened up new perspectives in a wide spectrum of applications. 
These range from the realization of a global-scale quantum communication network, the Quantum Internet \cite{kimble2008quantum,wehner2018quantum}, to the simulation of quantum systems \cite{georgescu2014quantum}, to quantum computing \cite{nielsen2001quantum}. 
In particular, the interest towards the last field has been fueled by some milestone discoveries, such as Shor’s factorization and Grover’s search algorithm \cite{shor1994algorithms,grover1996fast}, which have promised that quantum processors can outperform their classical counterparts. 
However, a clear advantage of quantum computation has been experimentally demonstrated only recently and on different computational tasks, boson sampling \cite{Aaronson2011a,zhong2020quantum,madsen2022quantum,Tillmann2013, broome_photonic_2013,crespi_integrated_2013} and random circuit sampling \cite{arute2019quantum}, which do not have clear practical applications.

\begin{figure*}[t]
    \includegraphics[width=1\textwidth]{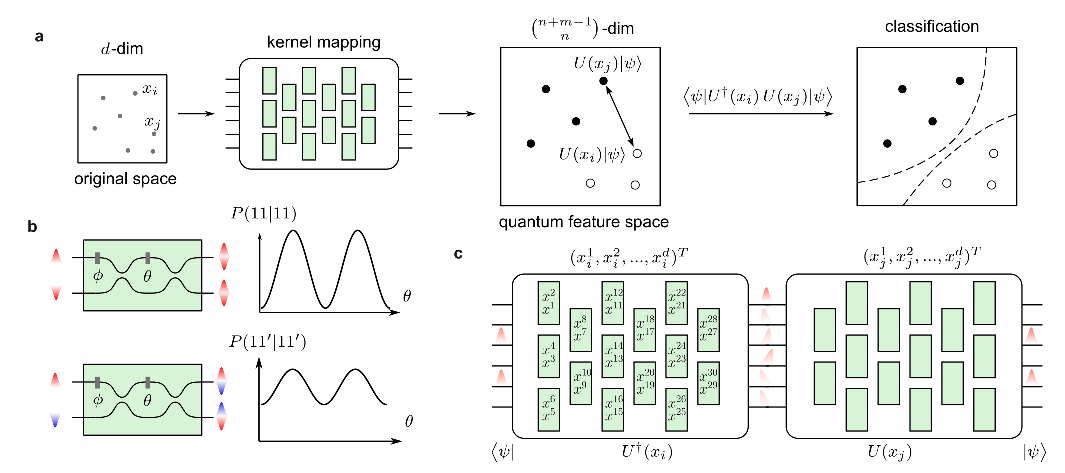}
    \caption{\textbf{Photonic quantum kernel estimation.}
    \textbf{a}.
    The photonic quantum kernel maps each data point $x_i$ to be classified from a $d$-dimensional space into a quantum state $|\Phi\rangle_i$, living in a Hilbert feature space.
    In detail, the classical data $x_i$ is encoded into a unitary evolution $U(x_i)$ applied on a fixed input state $\ket{\psi}$. This implies $|\Phi\rangle_i=U(x_i)|\psi\rangle$.
    After mapping all the data points in the dataset, from the inner pairwise products, we perform the classification finding the hyperplane best separating the classes, i.e. through a support vector machine (SVM), according to Eq.~\eqref{eq:classification}. 
    \textbf{b.} Pairs of indistinguishable photons and distinguishable photons show a different behaviour when injected in a Mach-Zehnder interferometer (MZI).
    Here, input states $11$ and $11'$ indicate, respectively, two indistinguishable and distinguishable photons being injected in the circuit and being detected at the output modes.
    \textbf{c.} Estimation of the inner product of two data points $x_i$ and $x_j$ by encoding them in two unitaries $U(x_i)$ and $U(x_j)$.
    The inner product $\langle\phi_j|\phi_i\rangle$ amounts to $\langle \psi|U^\dag(x_i)U(x_j)|\psi\rangle$. 
    This is equivalent to projecting the evolved state $U^\dag(x_j)U(x_i)|\psi\rangle$ onto $|\psi\rangle$. 
    Each box represents a programmable MZI with two free parameters (namely a beam splitter with tunable reflectivity and phase), as shown in \textbf{b}.
    }
    \label{fig:1}
\end{figure*}

Given these premises, \comment{our goal is to} investigate the tasks in which quantum computing can enhance the operation of classical computers \comment{for practically relevant tasks}. Moreover, the question is whether this can be achieved for problems that are now within the reach of state-of-art technology, where only noisy intermediate-scale quantum computers are available \cite{Preskill2018,brooks2019before}.
In this context, a flurry of interest has been devoted to the open question of whether the new paradigm of quantum computing can have an impact on machine learning \cite{biamonte2017quantum,wittek2014quantum,dunjko2018machine}, which has revolutionized classical computation, granting new possibilities and changing our everyday lives, from email filtering to artificial intelligence.
The two main directions that have been investigated until now are, on one side, whether quantum computation could improve the efficiency of the learning process, allowing us to find better optima with the need of a lower number of inquiries \cite{neven2009training,rebentrost2014quantum,leifer2008quantum,Saggio} and, on the other, how quantum behaviours can enhance the expressivity of the input encoding, exploiting correlations between variables that are hard to reproduce through classical computation \cite{boixo2018characterizing,gan2022fock}. 

In this context, a straightforward application of quantum computing on kernel models has become evident.
Kernel methods are widely used tools in machine learning \cite{shawe2004kernel,hofmann2008kernel}, that base their functioning on the fact that patterns for data points, which are hard to recognize in their original space, can become easy to identify once nonlinearly mapped to a \textit{feature space}. 
Once the suitable mapping is performed, it is possible to identify the hyperplane which best separates the classes of feature data points, through a support vector machine \cite{cortes1995support} (SVM), according to the inner product of the mapped data. 
Let us note that the only part of the model that is trained is the SVM, which is efficient, once the inner products are available.
Hence, an interesting question is whether using a quantum apparatus to perform the data mapping and evaluate the inner products can enhance the performance, benefitting from the quantum feature maps of the evolution of quantum systems and outsourcing the hardest part of the computation to the quantum hardware. 
This question was theoretically answered in the affirmative by \cite{Kolosova2020}, although the implementation of the proposed task is out of reach for state-of-the-art technologies. 
Moreover, a risk that one encounters in quantum kernel estimation is that, once the feature space is too large, points are mapped into orthogonal states, resulting in an ineffective classification. 
Hence, a moderately-sized quantum feature space can prove more suitable, to preserve the similarity among data belonging to the same class.

\begin{figure*}[t]
    \centering
    \includegraphics{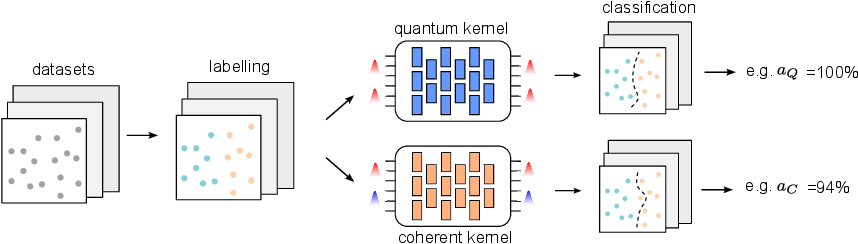}
    \caption{\textbf{Classification tasks for photonic kernel methods.} 
    The datasets are randomly generated and consist in $d$- dimensional vectors, with entries between 0 and 1. Then, we randomly assign labels to each point as belonging to class +1 or -1 and we test the ability of our photonic kernels, displaying and not displaying quantum interference (respectively indicated as \textit{quantum kernel} and \textit{coherent kernel}), to correctly classify the data.
    This is quantified by the accuracy of our models, which we indicate as $a_Q$ and $a_C$.
    }
    \label{fig:2}
\end{figure*}

\comment{
In this work, }
we experimentally demonstrate a quantum kernel estimation, where feature data points are evaluated through the unitary evolution of two-boson Fock states (see Fig.~\ref{fig:1}). 
Such encoding, even for relatively small dimensions, provides enough of a non-linearity to achieve high classification accuracies.
Furthermore, we show that for given tasks, this algorithm leads to an enhancement in the performance of quantum kernels with respect to their classical counterparts. 
These tasks are selected by maximizing the so-called \textit{geometric difference}, which measures the separation in performance between a pair of kernels \cite{Huang2021c}.
In particular, we separate between quantum and coherent kernels, 
that is, photonic kernels that do or do not exhibit quantum interference.

To experimentally demonstrate this method, we exploit a photonic platform based on an integrated photonic processor \cite{pentangelo_high-fidelity_2024} where we inject two-boson Fock states to map the data to be classified (see Fig.~\ref{fig:1}a). 
To estimate quantum and coherent kernels, we inject indistinguishable and distinguishable photons, respectively.
This photonic platform is particularly suitable for this task, as it allows us to encode and manipulate our input data with high fidelity. 

To benchmark our enhanced performance, we compare classification accuracies between photonic kernels and state-of-the-art classical computational kernels. 
These include the standard gaussian kernels~\cite{shawe2004kernel, hofmann2008kernel}, as well as the recently introduced neural tangent kernels~\cite{JacotEtAl18}, which simulate gradient descent over infinitely-wide neural networks.
Our results show that photonic kernels outperform classical methods 
and that the accuracies are further enhanced in kernels displaying quantum interference.

\section*{Photonic quantum kernel estimation}

A kernel method relies on a function that maps $N$ input data points $x_i$, on which we wish to perform binary classification, from a space $\mathcal{X}\subseteq \mathbb{R}^d$ into a feature space $\mathcal{H}$. 
Here, $d$ is the dimension of each data point.
This is done through a feature map $\Phi: \mathcal{X} \rightarrow \mathcal{H}$. 
Then, a SVM can be used to produce a \emph{prediction} function $f_K : \mathcal{X} \to \mathbb{R}$ as $f_K(x) = \sum_i \alpha_i K(x, x_i)$, where these $\alpha_i$ coefficients are obtained by solving a linear optimization problem. 
The inputs of the optimization are the labels $y$ and the matrix obtained by computing the pairwise distances between data points is 
\comment{
$K_{i, j} = K(x_i, x_j)=\abs{\braket{\Phi(x_i)}{\Phi(x_j)}}^2$, 
}
the so-called \emph{Gram matrix} (see Supplementary Note 1 for further information).

In this work, we implement a quantum version of the kernel method, in which the aforementioned pairwise distances between data points, which belong to a class $y$ taking values $+1$ or $-1$, are estimated by sampling from the output probability distribution arising from the unitary evolution of a Fock input state. 
This process is depicted in Fig.~\ref{fig:1}a. 
Therefore, our feature map plugs the data that needs to be classified into the free parameters defining a unitary evolution applied to a fixed Fock state of dimension $m$ and whose sum of occupational numbers is $n$: $x \mapsto |\Phi(x)\rangle= U_x |\psi\rangle$.
Here, $\ket{\psi}$ is the encoding state which is free to choose.
Then, as shown in Fig.~\ref{fig:1}c, the pairwise inner products of the feature points are experimentally evaluated, as $|\langle \psi|U(x_i)^\dag U(x_j)|\psi\rangle|^2$. 
Such unitaries can be effectively implemented by a programmable photonic circuit consisting of an array of Mach-Zehnder interferometers (MZIs)\cite{Clements2016}.
Hence, the dimension of the feature Hilbert space $\mathcal{H}$ will be $\binom{n + m - 1}{n}$. At this point, the SVM finds the hyperplane separating the training data points through the aforementioned optimization process \cite{boser1992training, vapnik1999nature} and,
the binary classification of unknown points $x$ is given by the following relation: 
\begin{equation}
    y=\mathrm{sign}\left(\sum_{i=1}^N \alpha_i y_i K(x, x_i)\right)
    \label{eq:classification}
\end{equation}
where $\alpha_i$ are 
the coefficients optimized in the training process and $y_i$ is the class of the i-th point in the training. 
This model is defined \textit{implicitly}, as the labels are assigned by weighted inner products of the encoded data points \cite{schuld2021supervised, Lloyd2020, kubler_inductive_2021,Bartkiewicz2020, Huang2021a, kusumoto2021experimental, haug2021quantum, scholkopf2002learning, jerbi2023quantum}.

\begin{figure*}[t]
    \centering
    \includegraphics[width=178mm]{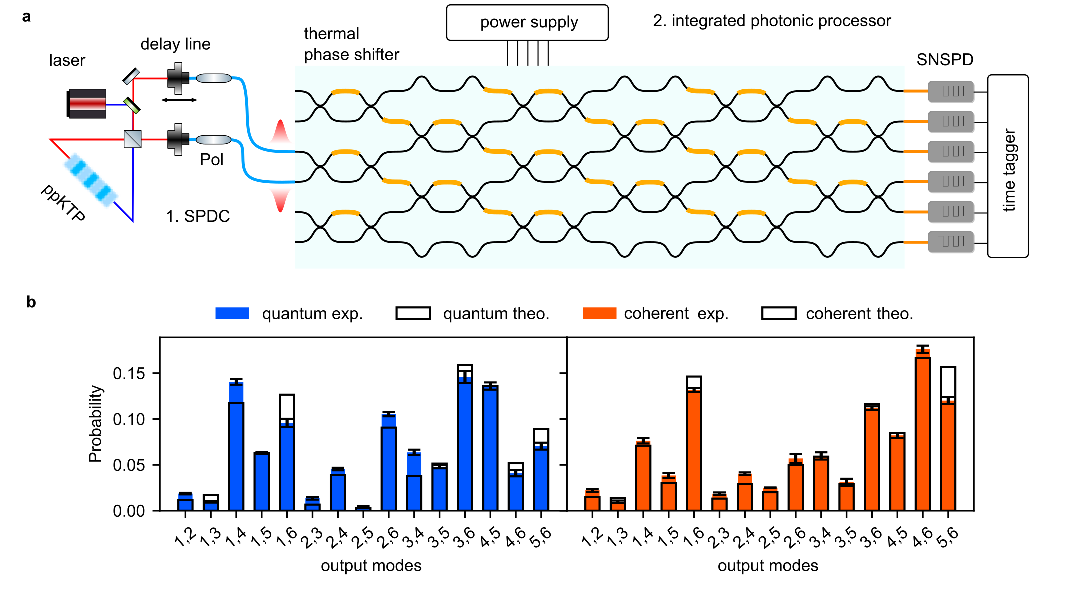}
    \caption{\textbf{Implementation of photonic quantum kernel estimation.} 
    \textbf{a.} Experimental setup consisting of two parts, the off-chip single photon source and the programmable integrated photonic processor.
    The frequency degenerate photons are generated by a type-II spontaneous parametric down-conversion source.
    Afterwards, the two photons are made indistinguishable in their polarization and arrival time. 
    Then, we inject these photons in two modes of an integrated photonic processor with six input/output modes \cite{pentangelo_high-fidelity_2024}. 
    Detection is performed by superconducting nanowire single-photon detectors (SNSPDs). 
    The degree of indistinguishability can then be tuned through a delay line, changing their relative temporal delay. 
    \textbf{b.} Probability distribution of photon detection events.
    We show two instances of the experimental photon detection probability, compared to the theoretical calculation.
    The quantum and coherent kernel measurements are obtained respectively by injecting two indistinguishable and distinguishable photons into the third and fourth modes of the circuits, i.e. $|0,0,1,1,0,0\rangle$. 
    The x axis shows all the circuit channels which output two photons simultaneously. 
    Thus, all 15 possible photon detection configurations are accessible.  
    }
    \label{fig:3}
\end{figure*}

If the Fock state contains indistinguishable bosons, they will exhibit quantum interference, as shown in Fig.~\ref{fig:1}b. 
In this case, the output probability distribution
is given by the permanents of sub-matrices of the matrix representing the unitary evolution of the input \cite{scheel2004permanents}. 
More specifically, considering an input configuration $s$, the probability of detecting the output configuration $t$ is given by ${ | \mathrm{Per} {U}_{s, t} |^2} / {\Pi_i^m s_i!\Pi_i^m t_i!}$. 
Here, $\mathrm{Per}(\cdot)$ denotes the permanent matrix operation, $s_i$ and $t_i$ are the occupational numbers at the $i$-th mode and $U_{s, t}$ is the sub-matrix obtained by selecting the rows/columns corresponding to the occupied modes of the input/output Fock states. 
On the other hand, if the bosons are distinguishable, they will not exhibit quantum interference.
In this case the probability will amount to ${\mathrm{Per}|U_{s,t}|^2} / {\Pi_i^m s_i!\Pi_i^m t_i!}$.

\comment{
In the following, we will refer to a kernel implemented with indistinguishable bosons as a \textit{quantum kernel},
\begin{equation}
    K_Q(x_i, x_j) = | \mathrm{Per} U_{\psi}(x_i, x_j) |^2 / N'
\end{equation}
and with distinguishable ones as a \textit{coherent kernel}, 

\begin{equation}
    K_C(x_i, x_j) = \mathrm{Per} |U_{\psi}(x_i, x_j)|^2  / N'
\end{equation}

Here, $U_{\psi}(x_i, x_j)$ is the matrix defined by data points $x_i, x_j$ and selected encoding state $\psi$. $N'$ is the coefficiency related to $\psi$.
As long as there is at most one photon in each mode of $\psi$, $N'=1$.

}

\section*{Classification task}\label{dataset-generation}

To select a classification task that would benefit from the described model, we use a quantifier called the \textit{geometric difference} \cite{Huang2021c}.
Given a set of data points $\{{x_i} | {x_i}\in \mathcal{X} \}$ without any labels, the geometric difference provides the binary labels $\{y_i\}$ that maximise the expected difference in prediction error between two kernels, 
$K_Q$ and $K_C$, as depicted in Fig.~\ref{fig:2}. 
Therefore, we obtain this optimal labelling by solving the following minimisation problem:
\begin{equation}
y^\star = \mathrm{arg}~\min \limits_{y \in \mathbb{R}^d} \left( \frac{s_{K_Q}(y)}{s_{K_C}(y)} \right)
\end{equation}
where $s_K(y) = y^T K y$ is the model complexity of the pair $K$ and $y$, i.e. the number of features that the model needs to make accurate predictions (see the Supplementary Note 2).
To saturate the following inequality \cite{Huang2021c}:
\begin{equation}
    \exists y \quad \cdot \quad s_{K_C}(y) \quad \leq  \quad g_{CQ}^2 \,s_{K_Q}(y) 
\end{equation}
we take $y^TK_C^{-1}y = g_{CQ}^2 y^TK_Q^{-1}y$, and obtain the relation
\begin{equation}
g_{CQ} = \sqrt{\norm{\sqrt{K_Q}\left(K_C\right)^{-1}\sqrt{K_Q}}_\infty}
\label{geom_diff}
\end{equation}
where $\norm{\cdot}_\infty$ denotes the spectral norm and $g_{CQ}$ denotes \textit{geometric difference}. 

We can now use Eq.~\eqref{geom_diff} to generate the classification task that,
given two kernels $K_Q, K_C$ and a set of data points $\{x_i\}$,
produces the labels $\{y_i\}$ that maximise the difference in prediction error bound. This can be done through the following procedure: 
(i) evaluate the Gram matrices $K_Q$ and $K_C$ over a set of non-labelled data points $\{x_i\}$; 
(ii) compute the positive definite matrix $M = \sqrt{K_Q}\left(K_C\right)^{-1}\sqrt{K_Q}$; 
(iii) compute the eigenvalues and eigenvectors of $M$ by spectral decomposition; 
(iv) find the maximum eigenvalue $g$ and its corresponding eigenvector $v$; 
(v) assign the labels $y = \sqrt{K_Q} v$. 
From a practical point of view, we start with the two aforementioned kernels, $K_C$ and $K_Q$, and then, by maximizing the geometric difference, we find the tasks for which the latter brings an enhanced accuracy of the classification. For more details regarding the algorithm to define the classification task, see the Supplementary Note 4. 
Let us note that the implemented tasks constitute instances of problems that can be naturally implemented with high accuracy on our quantum platform. 
As such they constitute a first stepping stone towards the identification of practical tasks for which quantum machine learning can enhance the performance of classical models.

\begin{figure*}[t]
    \centering
    \includegraphics[scale=0.85]{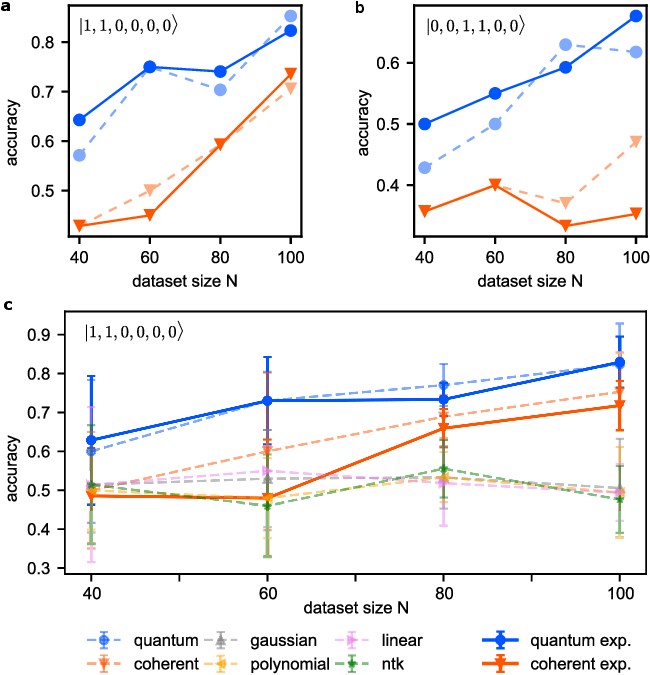}
    \caption{\textbf{Experimental classification accuracies.} 
    \textbf{a-b.} We tested our method on datasets of different sizes (40, 60, 80, 100) and for two different input states ($|1,1,0,0,0,0\rangle$ and $|0,0,1,1,0,0\rangle$) respectively. 
    For each dataset, $2/3$ of the data points were used for training the support vector machine (SVM) and $1/3$ for test. 
    \textbf{c.} The average classification accuracies on 5 different sets for the quantum kernel (blue) and the coherent (orange) kernel, 
    along with the following other computational kernels: gaussian (grey), ntk (green), polynomial (yellow) and linear (purple). 
    The dashed line indicates the results of numerical simulations, while the solid ones are the experimental results. 
    The error bar shows the standard deviation of the classification accuracies on 5 datasets for all the kernels.
    }
    \label{fig:4}
\end{figure*}

\section*{Experiment}

Our experimental setup consists of two parts, a single-photon source generating the input states and a programmable integrated photonic processor depicted in Fig.~\ref{fig:3}a.
First, to generate the input state, we use a type II spontaneous parametric down-conversion source, which generates frequency degenerate single-photon pairs at 1546~nm in a periodically poled K-titanyl phosphate crystal. The two photons are then made indistinguishable in their polarization and arrival time, respectively,  via wave retarders and a delay line, which we also use to tune the degree of indistinguishability of the generated photons. 

For the implementation of photonic kernels, which map our input data to a feature space, we require an apparatus able to perform arbitrary unitary transformations on a given input state. 
As mentioned before, our feature map sends each data point $x_i$ onto the state resulting from the evolution $U(x_i)$ of a fixed input Fock state $|\psi\rangle$. 
Then, for the application of the SVM, which finds the best hyperplane separating the data, we need to evaluate the inner products between all of the points $x_i, x_j$ in the feature space, which amounts to $\langle \psi|U(x_i)^\dag U(x_j)|\psi\rangle$. 
This implies that, if we take $|\psi\rangle$ as a Fock state of $n$ photons over $m$ modes, the inner product $\braket{ \Phi(x_i) }{ \Phi(x_j) }$ is given by projecting the evolved state $U(x_i)^\dag U(x_j)\ket{\psi}$ onto $\ket{\psi}$. 

To this aim, we employ an integrated photonic processor~\cite{pentangelo_high-fidelity_2024} on a borosilicate glass substrate, in which optical waveguides are inscribed through femtosecond laser writing~\cite{davis1996writing, osellame_femtosecond_2012, corrielli_femtosecond_2021}. 
The circuit features six input/output modes \cite{Clements2016}, as depicted in Fig.~\ref{fig:3}a, where each interferometer is equipped with two thermal phase shifters \cite{ceccarelli_low_2020}, to provide tunable reflectivity and phase. Such arrangement allows us to perform any unitary transformation on the input photon states. Given this property, our device is also referred to as a universal photonic processor. Design, fabrication and calibration of the integrated photonic circuit are described in \cite{pentangelo_high-fidelity_2024}.

Specifically, the data were encoded in the values of the phase shifts, as follows: $x_i=(x_i^1, x_i^2, ..., x_i^{30}) \rightarrow \theta_i=(2 \pi x_i^1, 2 \pi x_i^2, ..., 2 \pi x_i^{30})$, where $\theta_i$ are the phase shifts introduced by the phase shifters of a universal interferometer. \comment{Let us note that this encoding has the remarkable advantage that no extra processing is required on the input data, as they are directly plugged into the optical circuit parameters. Furthermore,} in principle, we would need a sequence of two of such circuits (as in the scheme of Fig.~\ref{fig:1}c), to first implement $U^\dag(x_i)$ and then $U(x_j)$ on our inputs. However, in our implementation, we adopt only one universal circuit and directly implement the unitary corresponding to the product $U(x_i)^\dag U(x_j)$. 
This reduces the experimental complexity and 
the circuit propagation losses.

At the output, detection is performed by superconducting nanowire single-photon detectors (SNSPDs), where we post-select the output events to those featuring two detector clicks \comment{(coincidence counts, CC)}, i.e. \textit{collision-free} events (see Supplementary Note 3). 
\comment{
Thus, the elements of the Gram matrix of a given kernel can be estimated from the coincidence counting 
$K(x_i, x_j) = \mathrm{CC}^{ij}_{\psi} / \sum_{1\leq l < m \leq 6} \mathrm{CC}^{ij}_{lm}$.
Here $\mathrm{CC}^{ij}_{lm}$ is the number of registered coincidence counts between channels $l$ and $m$, when the implemented unitary is $U^\dag(x_j)U(x_i)$ and $\psi$ indicates the occupied modes of input state $|\psi\rangle$.
}
To test the role of quantum interference in the accuracy of the classification, we tune the indistinguishability of the two photons by changing their relative temporal delay.
An instance of the probability distribution of the same unitary is shown in Fig.~\ref{fig:3}b.
The optimal classification task is chosen for each data set according to the algorithm explained in the previous section. 

\section*{Results}

We test the performance of two photonic kernels in several different configurations. 
Firstly, we consider two different inputs, $\ket{1,1,0,0,0,0}$ and $\ket{0,0,1,1,0,0}$. 
This amounts to either injecting the photons into the first two modes or the central two modes. 
Second, we are able to tune the indistinguishability to implement the quantum kernel and the coherent kernel as aforementioned.
During the whole measurement, the maximal achieved indistinguishability between the photons is $0.9720\pm0.0044$, estimated through on-chip Hong-Ou-Mandel interference \cite{hong1987measurement}. 

For both input states, we fix the encoding of each data point and vary datasets with four different sizes: 40, 60, 80 and 100.
We use the setup depicted in Fig.\ref{fig:3}a to implement all pairwise products between the unitaries $U(x_i)^\dag U(x_j)$. 
Hence, $|\langle \psi |U(x_i)^\dag U(x_j)|\psi\rangle|^2$ is given by the probability of detecting the photons on the same modes from which they were injected. 
\comment{The rate of total post-selected coincidence counts amounts to 10~kHz and the measured probability distribution was averaged over 5~s for each unitary configuration.}

For each size $N$, we perform $N(N-1)/2$ unitaries to compute the inner products. 
The distance between the unitaries experimentally realized and the target ones can be estimated as $\sum_i \sqrt{P_i^\mathrm{theo} \cdot P_i^\mathrm{exp}}$, where $P_i^\mathrm{exp}$ is the experimental detection frequency for the $i$-th output configuration, while $P_i^\mathrm{theo}$ is the one estimated based on the theory\cite{Scheel2004}.
The mean fidelity of all datasets is 0.9816$\pm$0.0148 and 0.9934$\pm$0.0048, for the quantum kernel and coherent kernel respectively.
For each dataset, we use $2/3$ of the data points for training of the SVM, which yields the coefficients mentioned in \ref{eq:classification}.
The remaining $1/3$ as the test dataset can be used to predict the classification accuracy.
Accuracy is defined as the percentage of correctly classified points out of the total size of the test set.
Let us note that values lower than 0.5 indicate that the model was not able to learn the features of the training set and generalize to unknown data.

In Fig.~\ref{fig:4}a and b, we show the test accuracies obtained by injecting two input states for four different dataset sizes, where the quantum kernel performs significantly better than the coherent kernel at both experiments.
In Fig.~\ref{fig:4}c, we report the average test accuracy obtained for five different datasets with the same size, varying the dataset sizes from 40 to 100 as well.
Moreover, the results obtained with the quantum kernel (blue) and the coherent kernel (orange) are compared with the following numerical kernels: neural tangent kernel (green)~\cite{JacotEtAl18}, gaussian (grey), polynomial (yellow) and linear (purple)~\cite{shawe2004kernel, hofmann2008kernel}. 
Here, the neural tangent kernel (ntk) adopts a infinite-width neural network to classify the data optimized by gradient descent. 
See the Supplementary Note 5 for more details.
The dashed lines indicate the results of numerical simulations, while the solid lines indicate experimental results.
Although the task is built only comparing the performance of the kernels based on indistinguishable and distinguishable photons, the both obtained accurracies are higher than other classical kernels.

\section*{Discussion}

In this work, we show the first experimental demonstration of quantum kernel estimation, based on the unitary evolution of Fock states through an integrated photonic processor. 
We map data into a feature space through the evolution of a fixed two-photon input state over six modes. 
To achieve this, we adopt an quantum processor realized by femtosecond laser writing in a borosilicate glass substrate \cite{pentangelo_high-fidelity_2024}. 
The sampled output distribution is then fed into an SVM, performing the classification. 
Note that, although our apparatus only features tunable phase shifters and beam splitters, such encoding produces a sufficient non-linearity to achieve high classification accuracy of non-linearly separable datasets. 
This constitutes a difference of our method from alternative platforms, where entangling gates are typically needed \cite{havlivcek2019supervised, glick_covariant_2024, Huang2021c}.
Furthermore, in our case, it is not necessary to increase the dimension of the feature Hilbert space to achieve a good accuracy. 
This circumvents the typical difficulty of quantum kernels whereby all data points are encoded in orthogonal states, leading to ineffective classification 
\comment{
\cite{kublerInductiveBiasQuantum2021, thanasilpExponentialConcentrationQuantum2024}. Moreover, the fact that our model is effective for small dimensions is a crucial feature,  since we require an approximation of the full probability distribution deriving from the evolution of our input state. Hence this study is relevant for medium-sized problems, because reaching high dimensions would imply the input/output combinations to grow exponentially, along with the number of required experimental shots to reach arbitrary accuracy.
}

The task we implement is designed by assigning binary labels to randomly generated data points, which we encode in the phase shifts of an optical circuit. 
This is done exploiting the so-called \textit{geometric difference}, that selects the task for which the presence of quantum interference yields a better classification accuracy compared to cases where no interference is displayed. 
Although, the geometric difference compares the performance of a pair of kernels (in our case kernels implemented with indistinguishable/distinguishable bosons), for the selected tasks, both photonic kernels performed significantly better than commonly used kernels, not only the gaussian, polyonomial and linear ones, but also the neural tangent kernel.
Our results indicate that a photonic kernel estimation can enhance the performance, even for medium-size problems, reachable by current quantum technologies. 
Moreover, using distinguishable bosons to have a (smaller) performance enhancement represents an intriguing possibility, as it can prove convenient to reduce the impact of photon losses on the experimental time required to collect significant statistics.

\comment{Let us highlight that, although the interference of two photons in a six-mode unitary matrix (or more in general for medium-sized problems) can be estimated by classical computers, this does not affect the features of our protocol. Firstly, because the approximation of permanents through classical algorithms, e.g. Gurvits one \cite{gurvits2005complexity} has a slightly worse performance when compared to the direct sampling from an optical circuit, which naturally implements the studied kernel. In particular, the first scales as $O(n^2/e^2)$, whereas $n$ is the number of photons and $e$ is the required precision, while the direct sampling as $O(1/e^2)$ \cite{aaronson2012generalizing}.}
This is also the reason for which we do not use our photonic platform to reproduce classical kernels, as in \cite{bartkiewicz2020experimental}, being given by the natural evolution of bosons through a quantum circuit. Moreover, this protocol sheds light on alternative computational models, exploiting optical computation. This may be of particular importance when considering difficulties related to energy consumption, as it has been proved that partially optical networks can reduce energy requirements with respect to electronic ones \cite{hamerly2019large}. 

Despite being overshadowed by deep neural networks, kernels are still widely used in a large number of tasks, due to their simplicity, and ability to learn from small datasets \cite{lee2020finite, radhakrishnan2023transfer}.
Indeed, they have seen a recent revival in classical machine learning
with the introduction of neural tangent kernels ~\cite{JacotEtAl18} and their use in the study of state-of-the-art neural network architectures such as transformers~\cite{transformer}.
Another recent trend consists in merging neural networks and kernels, where notable examples are attention modules in natural language processing, and Hopfield layers \cite{ramsauer2020hopfield}. 

Our method can find a wide range of promising near-term applications in tasks such as information retrieval, natural language processing and medical image classification~\cite{WangEtAl21,YuEtAl16a,LorenzEtAl21,LandmanEtAl22}, where kernels have been proposed as a keystone~\cite{SchuldPetruccione21a}.
Our experimental results also open the door to hybrid methods where photonic processors are used to enhance the performance of standard machine learning methods.
They also bring forward investigations on the non-linearities that can be achieved through photonic platforms \cite{denis2022photonic, Spagnolo}, in particular, for neuromorphic computation models, such as \textit{reservoir computing} \cite{govia2022nonlinear, innocenti2023potential}.  In addition, we envisage the combination of this kind of non-linearity with those brought by the implementation of feedback loops, as in the case of quantum memristor \cite{Spagnolo} and the exploitation of quantum interference in the implementation of feature maps.


\section{Acknowledgements}
The authors would like to thank Anna Pearson for her active involvement in the first stages of this project. 
This research was funded in whole or in part by the Austrian Science Fund (FWF)[10.55776/ESP205] (PREQUrSOR), [10.55776/F71] (BeyondC), [10.55776/FG5] (Research Group 5) and [10.55776/I6002] (PhoMemtor). For open access purposes, the author has applied a CC BY public copyright license to any author accepted manuscript version arising from this submission.
This project has received funding from the European Union’s Horizon 2020 research and innovation programme under grant agreement no. 899368 (EPIQUS), the Marie Skłodowska-Curie grant agreement No 956071 (AppQInfo) and grant agreement no. 101017733 (QuantERA II Programme, project Phomemtor). Views and opinions expressed are however those of the author(s) only and do not necessarily reflect those of the European Union or the European Research Council. Neither the European Union nor the granting authority can be held responsible for them. 
The financial support by the Austrian Federal Ministry of Labour and Economy, the National Foundation for Research, Technology and Development and the Christian Doppler Research Association is gratefully acknowledged.
G.d.F would like to thank Konstantinos Meichanetzidis and Tommaso Salvatori for insightful discussions on the project. 
The integrated photonic processor was partially fabricated at PoliFAB, the micro- and nanofabrication facility of Politecnico di Milano (\href{https://www.polifab.polimi.it/}{https://www.polifab.polimi.it/}). C.P., F.C. and R.O. wish to thank the PoliFAB staff for the valuable technical support. 
R.O.  acknowledges financial support by ICSC – National Research Center in High Performance Computing, Big Data and Quantum Computing, funded by European Union– NextGenerationEU.

\section{Author contributions}
Z.Y. and I.A. designed and conducted the experiment and G.d.F., D.B. and A.T. developed the theory and algorithm. C.P., S.P., A.C. and F.C. conducted the design, fabrication and calibration of the integrated photonic processor. Z.Y., I.A. and G.d.F. wrote the first draft of the manuscript. 
R.O., B.C. and P.W. supervised the whole project.
All authors discussed the results and reviewed the manuscript.

\section{Competing interests}
F.C. and R.O. are cofounders of the company Ephos. 
The authors declare that they have no other competing interests.

\newpage

\section{Methods}

The two photon input states are generated by a type-II spontaneous parametric down-conversion source, which generates frequency degenerate single-photon pairs at 1546~nm via a 3-cm long periodically poled K-titanyl phosphate (ppKTP) crystal. 
Afterwards, the two photons are made indistinguishable in their polarization, which is rotated through paddles, and in arrival time, through a delay line, which we use also to tune the degree of indistinguishability of the generated photons.

Then, we inject these photons into two selected modes of an integrated photonic processor with six input/output modes\cite{pentangelo_high-fidelity_2024}. 
This circuit features 27 thermal phase shifters and its architecture follows the rectangular scheme presented in \cite{Clements2016}, to implement arbitrary unitary evolution on any input Fock state. 
Compared with the original architecture, the first three external phase shifters are omitted, as they will not effect the probability distribution.
To apply accurate phases independently, each channel is supplied by a current source to avoid electrical crosstalk (4$\times$Qontrol q8iv, 16bit DAC).
In the end, detection is performed by superconducting nanowire single-photon detectors (SNSPDs) housed in a 1K cryostat. 
We post-select the detected events to the cases in which two detectors click simultaneously in a temporal window of 1~ns.
A time tagger with a 15.63~ps resolution is used to process the real-time coincidence counting for all 15 post-selection patterns.
\comment{
The total coincidence counting is about 10~kHz, which varies due to the pump laser and the detection efficiency.
For each unitary configuration, we integrate 5~s to estimate the probability distribution over 15 coincidence patterns.
}

\section{Data availability}
The data supporting this study’s findings are available from the project page (\href{https://github.com/dapingQ/PhoQuKs}{https://github.com/dapingQ/PhoQuKs}), containing detailed explanations of all the datasets.

\section{Code availability}
The code scripts analyzing the study is available from the project page (\href{https://github.com/dapingQ/PhoQuKs}{https://github.com/dapingQ/PhoQuKs}).

\end{document}


\title{Supplementary Information
\\
Experimental quantum-enhanced kernels on a photonic processor}

\maketitle

\section{Support vector machines and kernel methods}\label{SVM}

For a dataset consisting of $N$ data points with a $\pm1$ label, $D = \set{(x_i, y_i)}_{i=1}^N$, each data point is a $d$-length vector, 
$x_i = (x_i^1, x_i^2 ..., x_i^d)^T$, where $x_i \in \mathcal{X} \subseteq \mathbb{R}^d$ and $x_i^j$ is the $j$-th feature of the input vector $x_i$. 
$y_i \in \mathcal{Y} = \set{+1, -1}$ is the binary label corresponding to the $x_i$.
The classification task is to find a hyperplane $w\in \mathbb{R}^d$ and a bias parameter $b \in \mathbb{R}$, thus a linear \emph{support vector machine} (SVM) is defined to predict the label of unknown data points $x$.
\begin{equation}
f(x) = \mathrm{sign} (w\cdot x +b)    
\end{equation}
This is an optimization problem,
\begin{align}
    \min \quad& \frac{1}{2}||w||^2 \\
    \mathrm{s.t.} \quad& y_i(w \cdot x_i+b) -1 \geq 0, \quad i=1,2,...,N 
\end{align}
and it equals to solve the dual problem,
\begin{align}
    \min_\alpha \quad & \frac{1}{2} \sum_{i=1}^N \sum_{i=1}^N \alpha_i \alpha_j y_i y_j (x_i \cdot x_j) - \sum_{i=1}^N \alpha_i \\
    \mathrm{s.t.} \quad & \sum_{i=1}^N \alpha_i y_i = 0 \\
    & \alpha_i \geq 0, \quad i=1,2, ..., N
\end{align}
The solution $\{\alpha_i\}$ are obtained by solving a linear optimization problem, given by the inner products of all input vectors $x_i \cdot x_j$ and the labels $y_i$.

Instead, rather than the direct linear inner products, the kernel function is used to transform the input vector from the original input space to another inner product \emph{feature space}, $\Phi : \mathcal{X} \to \mathcal{H}$.
The \emph{Gram matrix} is a symmetric matrix where each element represents the pair-wise of inner product in a feature space $K_{ij} = \Phi(x_i)\cdot \Phi(x_j)$. The prediction function based on a kernel $K$ is given as
\begin{equation}
   f_K(x) = \sum_i^N \alpha_i K(x, x_i) 
\end{equation}

For some implicit kernel functions, it is easier to calculate the gram matrices than the kernel functions. For example, if we define the feature space as \emph{Fock space} $\mathcal{F}$, each original data point is mapped to a quantum state $x_i \mapsto \ket{\Phi(x_i)}=U(x_i)\ket{\psi}$ by a parameterized circuit $U(x_i)$ and Fock state $\ket{\psi}$. 

\comment{
Instead of performing quantum state tomography over every inner product, which is an expensive task considering the size of feature spaces, \cite{banchi_multiphoton_2018}, 
it is straightforward to measure the outcome probability gram matrix $K(x_i, x_j) = \abs{\braket{\Phi(x_i)| \Phi(x_j)}}^2 = \abs{\braket{\psi|U(x_i)^\dag U(x_j)|\psi}}^2$.
}
Thus, in our experiment, the task is to estimate each element of gram matrices for each dataset which is equivalent to estimate the outcome probability of  
In Figure~\ref{fig:gram}, we show an example of experimentally reconstructed Gram matrices, as well as the ones calculated by the theory.

\section{Geometric difference}\label{geo-diff}

\begin{figure}[!h]
    \centering
   \includegraphics{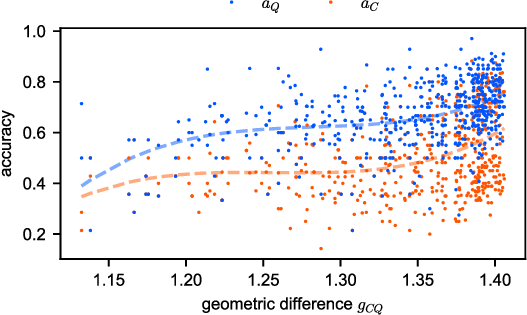}
   \caption{\textbf{Correlation between geometric difference and classification accuracy.} The blue dots indicate the numerical accuracy achieved on test datasets of several dimensions exploiting the quantum kernel, varying the unitary ansatz width, encoding state and dataset sizes. The red ones, instead, refer to the coherent kernel. The figure shows that\comment{, on average, the geometric difference is higher than 1, which implies that the quantum kernel has higher model complexity than the coherent one.}}
   \label{fig:geod}
\end{figure}

For a given kernel method and a dataset $D = \set{(x_i, y_i)}_{i=1}^N$, the \emph{model complexity} of the triple $(K, N, y)$ is given by:
\begin{equation}
    s_K(y) = \sum_i\sum_j(K^{-1})_{ij} y_i y_j = y^T K^{-1} y
\end{equation}
As shown in  \cite{Huang2021c}, 
this quantity can be used to bound the prediction error of a kernel method $K$ on the dataset:
\begin{equation}
    \mathbb{E} \vert  f_K(x) - f(x)  \vert \quad \leq \quad c \cdot \sqrt{\frac{s_K(y)}{N}}
\end{equation}
Here, $x \in \mathcal{X}$, $c$ is a constant and $N$ is the number of data points.

The concept \emph{geometric difference} is used to separate the complexity of two kernels.
In this work, we target two kernels $K_Q$ and $K_C$, which are the transform from original space to the Hilbert space given by the evolution of indistinguishable photons and the one given by distinguishable photons, respectively.
Therefore, the optimal labelling is attained by solving the optimization problem by minimizing the ratio of the model complexity $s_{K_Q}$ to $s_{K_C}$:
\begin{equation}
y^\star = \arg\min\limits_{y \in \mathbb{R}^N}  \frac{s_{K_Q}(y)}{s_{K_C}(y)}
\end{equation}
The solution is $y=\sqrt{K_Q} \vb{v}$, where $\vb{v}$ is the eigenvector of matrix $\sqrt{K_Q}(K^C)^{-1}\sqrt{K_Q} $ with the eigenvalue noted as $g_{CQ}^2$ \cite{Huang2021c}.
This solution holds the inequality:
\begin{equation}
    s_{K_C}(y) \quad \leq  \quad g_{CQ}^2 \,s_{K_Q}(y) 
\end{equation}
This inequality saturates when $g^2$ equals the spectral norm, as follows
\begin{equation}
    g_{CQ} = \sqrt{\norm{\sqrt{K_Q} (K_C)^{-1}\sqrt{K_Q}}_\infty}
\end{equation}

In practice, a regularization parameter $\lambda$ is used in the model complexity $s_K$, thus the geometric difference is alternated as 
\begin{equation}
    g_{CQ} = \sqrt{\norm{\sqrt{K_Q} (K_C+\lambda I)^{-1}\sqrt{K_Q}}_\infty}
\end{equation}
In the following simulation, we use $\lambda=0.02$ without loss of generality.

Based on this model, we perform simulations of randomly generated datasets, varying number of data points, dimension of unitaries and photon number, to explore the correlation between geometric difference and test accuracies, which we show in Figure~\ref{fig:geod}. As expected, there is a correlation between the enhancement in the accuracy and the value of the geometric difference. 
It is also noteworthy that, although the task was built only comparing the performance of these two kernels, the obtained accuracy is higher also than commonly used standard kernels.

\begin{figure}[!ht]
    \centering
   \includegraphics[width=100mm]{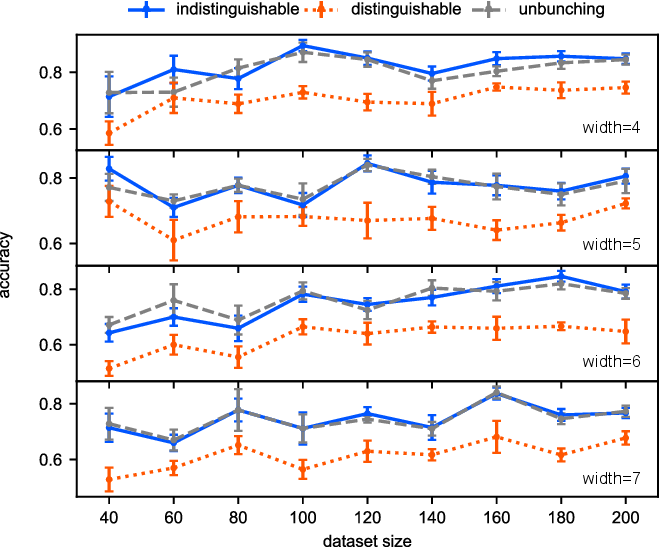}
   \caption{\textbf{Classification accuracies under approximation using unbunching photons} 
   Simulation of the classification accuracies using quantum kernels, both considering and not considering bunching events and coherent kernels, varying different circuit widths.}
   \label{fig:unbunching_plot}
\end{figure}

\section{Approximation of bosonic kernels with post selection}\label{app-unbunching}

The inner product in bosonic kernels $K_Q$ and $K_C$ can be measured experimentally by collecting the probability over all output photon states $\set{\phi_{m,n}}$, which corresponds to the normalization of coincidence counting for all possible configuration.
Here $m$ is the mode number of the quantum circuit $U(\theta)$ with parameters $\theta$ and $n$ is the photon number.
For example, for an input state $s$ and a output state $t$,  
\begin{equation}
    P_\theta(t|s) = \frac{\cc (t )}{\sum_{k \in \set{\phi_{m,n}} }{\cc(k )} }
\end{equation}
including both unbunching (collision-free) photon states $\mathbb{U}$ and bunching photon states $\mathbb{B}$, $\set{\phi_{m,n}}=\mathbb{U}\oplus\mathbb{B}$, which requires photon number resolving detectors for each output mode.
Since we are unable to detect those bunching events, 
a common solution to this issue is to only keep the statistics of unbunching events \cite{Tillmann2013}, 
when all photons output from the circuits are detected in separate modes.
We normalize the probabilities of unbunching states $\mathbb{U} \subsetneq  \set{\phi_{m, n}}$: 
\begin{equation}
    P_\theta (t \vert s) \approx P^\mathbb{U}_\theta (t \vert s) =\frac{\cc (t )}{\sum_{k \in \mathbb{U}}\cc(k)}
\end{equation}  \cite[Section 13]{Aaronson2011a}
As shown in  (\textit{7}) the unbunching probabilities are a good estimate of the ideal bosonic statistics when  $m >> n^2$.
We can use the these statistics to define the \emph{unbunching kernel}:
\begin{equation}
    K(x_i, x_j) \approx K_U(x_i, x_j) = P^\mathbb{U}_\theta (\psi \vert \psi) 
    = \frac{\cc_{\psi}}{\sum_{1\leq i < j \leq 6} \cc_{ij}}
\end{equation}
for some state $\psi \in \mathbb{U}$.
Under this approximation, the unbunching kernel is not a positive definite kernel since the estimated probability is not real inner products in a Hilbert space.
In practice however, the function can still be used in a support vector machine and our simulations show that the resulting leaning algorithm performs similarly to the bosonic kernel, see Fig \ref{fig:unbunching_plot}. 

In this case, the labeling is still optimized for the quantum kernel $K_Q$ with respect to the coherent kernel $K_C$. 
We observe that the unbunching kernel performs similarly to the case where all of the output events are considered and consistently better than classical particles. This indicates this approximation is good enough to reproduce the quantum kernel by detecting only the collision-free photons.

\section{Intermediate degrees of photon indistinguishability}

\begin{figure*}[!h]
    \centering
    \includegraphics{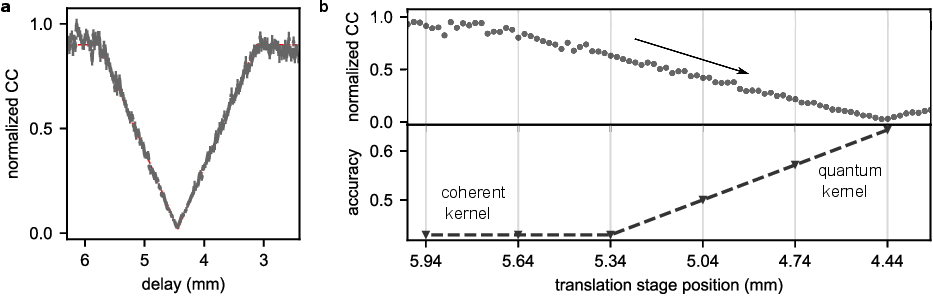}
    \caption{ \textbf{Classification test accuracy vs degree of photon indistinguishability.}
    \textbf{a}. On-chip Hong-Ou-Mandel interference. 
    The maximal visibility achieved amounts to 0.9720$\pm$0.0044.
    \textbf{b}. Transition from fully distinguishable to fully indistinguishable photon inputs. 
	}
    \label{fig:trans}
\end{figure*}

To test the role of photon indistinguishability, we change the relative temporal delay between the input photons, making them distinguishable in their arrival times. 
This implies that, for delays that are lower than the coherence time of photons, we are sending the following mixture in the chip: $r\ket{1,1,0,0,0,0}+(1-r)\ket{1,1',0,0,0,0}$, with $0 < r < 1$.
Here, $r$ is defined as the \textit{degree of indistinguishability} \cite{renema2018}.
We report this transition, with the corresponding test accuracies for a data set of 40 elements in Figure~\ref{fig:trans}.
These values are obtained for datasets of 40 elements, where $2/3$ are used as training and $1/3$ as test set, with input state $|0,0,1,1,0,0\rangle$.

\section{Numerical experiments}

\begin{figure}[!ht]
	\centering
	\includegraphics[width=1\textwidth]{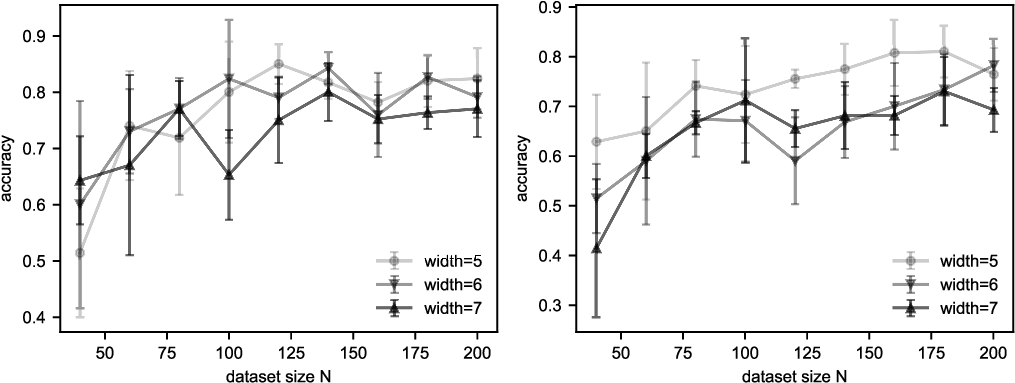}
	\hspace{0.065\textwidth} 
	\caption{
		\textbf{Accuracy of the quantum kernel $K_Q$ against the number of data points for several widths of the integrated circuit.} The two plots refer to two different input states, i.e. on the left, we consider the state $|1,1,0,0,0,0\rangle$ and, on the right, we consider $|0,0,1,1,0,0\rangle$.
		Each point is the mean accuracy on 5 iterations, each starting with randomly generated data.
		The standard error on the mean is shown by error bars.
		For each width, we observe a gradual increase in classification accuracy with number of data points.
		Typically, lower widths perform better for the same number of data points.
		We hypothesise that this is because the number of parameters available to the system increases with
		width, and so too many parameters on too little data leads to data being sparse in the space
		(i.e. all data points being approximately linearly independent so difficult to classify).
	}
	\label{fig:widths}
\end{figure}

\begin{figure}[!ht]
	\includegraphics{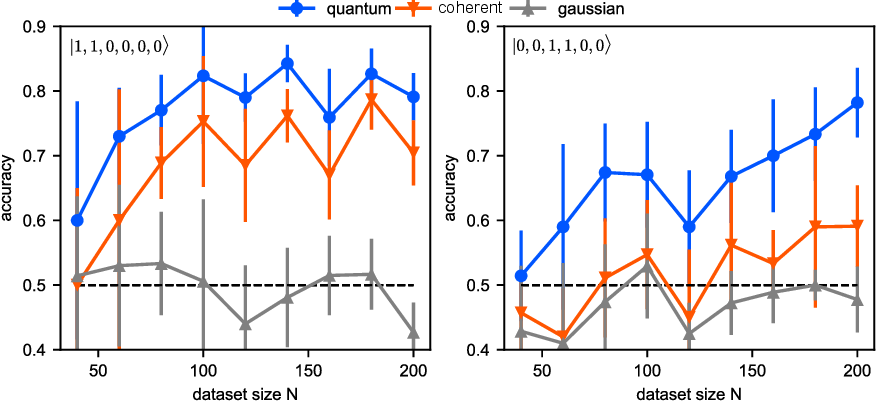}
	\caption{\textbf{Classification accuracies of photonic and Gaussian kernels.} We show the numerically estimated average test accuracies over 5 iterations for quantum kernel, coherent kernel and Gaussian kernels as the number of data points grow.
	The black dash lines indicate the random guessing accuracy of 50\%.
	The data for the quantum kernel comes directly from Figure \ref{fig:widths}
	-- the width that produced the highest accuracy is shown.
	The standard error on the mean is shown by error bars. 
	We observe separation between the classification accuracy of the quantum and coherent kernels.
	This is more prominent for the centred initial state (right-hand figure).
	Both perform considerably better than the Gaussian kernel on the same data,    which has an average accuracy of less than 60\%.}
	\label{fig:qcg_plot}
\end{figure}

\subsection{Dataset generation algorithm}
In the Algorithm \ref{algo}, we report the algorithm used for all simulation mentioned in this work.

\begin{algorithm}[H]
\caption{Dataset Generation}\label{algo}
    \begin{algorithmic}
    \State \textbf{input} circuit width $m$, depth $k$ of the chip, number of input photons $n$, initial state $s \in \Phi_{m, n}$, number of data points $N$
    \State generate $N$ random data points $\{x_i\}_{i=1}^N, x_i \in \mathbb{R}^d$
    \Comment{ Note  $d$ is proportional to $mk$} 
    \State $U(x_i) \gets \prod_{j,k\in C} \mathrm{{SU}_{MZI}}(x_i^{(j)}, x_i^{(k)})$
    \Comment{$C$ is the index in Clements coding}
    \State
    \Comment{$\mathrm{{SU}_{MZI}}$ is the matrix of Mach-Zehnder interferometer.}
    \While{ $ 0<i \leq N $ }
    \Comment{calculate the gram matrix}
    \While{ $ 0< j \leq i $ }
    \State  $K_Q(x_i, x_j) \gets |\mathrm{per}(U(x_i, x_j))|^2$
    \State  $K_C(x_i, x_j) \gets \mathrm{per}(|U(x_i, x_j)|^2)$
    \State $ j \gets j + 1$
    \EndWhile
    \State $ i \gets i + 1$
    \EndWhile
    \State $S \gets \sqrt{K_Q} (K_C+\lambda I)^{-1}\sqrt{K_Q}$ 
    \State $ \lambda \gets \max{ \mathrm{eigenvalue~of~} S }, \mathbf{v} \gets \max{\mathrm{eigenvector~of~} S}$
    \State $g_{CQ} \gets \sqrt{\lambda}$
    \State $y \gets \mathrm{sign}{(\mathbf{v})}$
    \Comment{apply sign function to get binary labels}
    \State \textbf{output} $D = \{(x_i, y_i)\}_{i=1}^N$.        
    \end{algorithmic}
\end{algorithm}

\subsection{Benchmark algorithm}

To benchmark the photonic kernels, we choose the following four kernels to compare the classification accuracies, gaussain kernel, polynomial kernel, liner kernel and neural tangent kernel.
First, for the gaussian kernel, the gram matrices are calculated as 
\begin{equation}
    K_G(x_i, x_j) = \exp(\gamma \vert \, x_j - x_i \, \vert^2)
\end{equation}
Here $\gamma$ is a hyper parameter and $x_i, x_j$ are two different data points.
Second, the polynomial kernel is defined as
\begin{equation}
    K_P(x_i, x_j) = (\gamma x_i \cdot x_j + r)^d 
\end{equation}
where there three hyper-parameters $\gamma r,d$. In both gaussian kernel and polynomial kernel, we run the grid search to find the optimal hyper-parameters.
Next, for the linear kernel, we take the simple linear inner product as
\begin{equation}
    K_L = x_i \cdot x_j
\end{equation}
For the above three kernels, the gram matrices are processed by a classical support vector machine to predict the data points in the test datasets.

Last but not least, to implement the neural tangent kernel, we adopt the open source program \cite{neuraltangents2020}
which trains an ensemble of infinite width neural networks using gradient descent.
The network consists two one input layer, two hidden layers and one output layer.
Without loss of generality, the two hidden layers both include 30 neurons, as the input
data points are 30 dimensional.
The sign function is applied on the output to identify the class.
In fact, the hyper parameters in this infinite work does not affect the accuracy as the labeling is very specific to separate the photonic kernels.

\subsection{Simulations}
We conduct the following numerical simulations to substantiate the feasibility of our method. The classification accuracy by quantum, coherent, gaussian, neural tangent, polynomial and linear kernels are labeled as $a_Q, a_C, a_G, a_N, a_P, a_L$ below.

\paragraph{Task1}
As a first simulation, we fix photon number $n$, input state $s$ and dataset size $N$, and get 3D plots of width $m$, depth $k$ and accuracy $\set{a_Q, a_C, a_G}$.
Results indicate that width and depth should be linearly related to maximise separation.
In subsequent experiments we used $m = k$. 
We used two types of initial states, Left state $\ket{\psi_L} = \ket{1,1,0,0,0,0}$ where photons are input of one side of the chip and Cent state $\ket{\psi_C} = \ket{0,0,1,1,0,0}$ where they are input in the middle.
\paragraph{Task 2}
In our second simulation, we fix photon number $n$ and the encoding state $s$, set $m = k$ and plot the quantum
accuracy $a_Q$ and coherent accuracy against the number of data points $N$. The results are given in Figure~\ref{fig:widths}.
We can observe this point experimentally by checking that $s_Q = g_{CQ}^2 s_C$.
When this is violated the gram matrix $K_Q$ is not invertible, there are too many
linear dependencies between the columns, i.e. too many data points in a space of dimension $\binom{n + m - 1}{n}$.
\paragraph{Task 3} 
Now we have a way of scaling width $m$, depth $k$ and number of data points $N$ together, by taking the best chip size $m = k$ for a given number of data points.
We plot the accuracies for quantum, coherent and gaussian kernels as the system size grows. The results are given in Figure \ref{fig:qcg_plot}.
\paragraph{Task 4} 
We also performed simulations to test unbunching kernels on the datasets resulting from separation between quantum and coherent gram matrices, see Figure \ref{fig:unbunching_plot}. 
Separating directly between unbunching and distinguishable kernels is often not possible.
This is because the separation method requires to compute the square root of the gram matrix, but the one obtained from unbunching statistics is often not completely positive.

\section{Experiment notes}

As described in the main text, the whole experiments were conducted to calculate each element of gramm matrices. 
Since the gram matrices are symmetric, $K_{ij}=K_{ji}$, thus for $N$ data points, we run $N(N-1)/2$ times to get the gramm matrices. 
In Figure 3\textbf{c}, for each dataset size varying from 40 to 100, we have 5 independent iterations, this indicates 53,300 running of unitary for both quantum and coherent kernels.
The total experiments lasted hundreds of hours.

\begin{figure*}[!ht]
	\centering
	\includegraphics{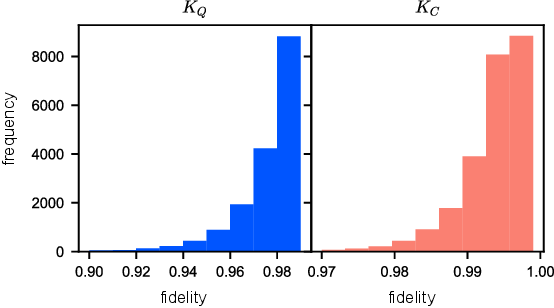}
	\caption{\textbf{Experimental probability fidelity of the unitaries}. The experimental fidelity for all data points performed by the quantum kernel (blue, left) and the coherent kernel (orange, right).
		.}
	\label{fig:fid}
\end{figure*}

\begin{figure*}[!h]
	\centering
	\includegraphics[width=5in]{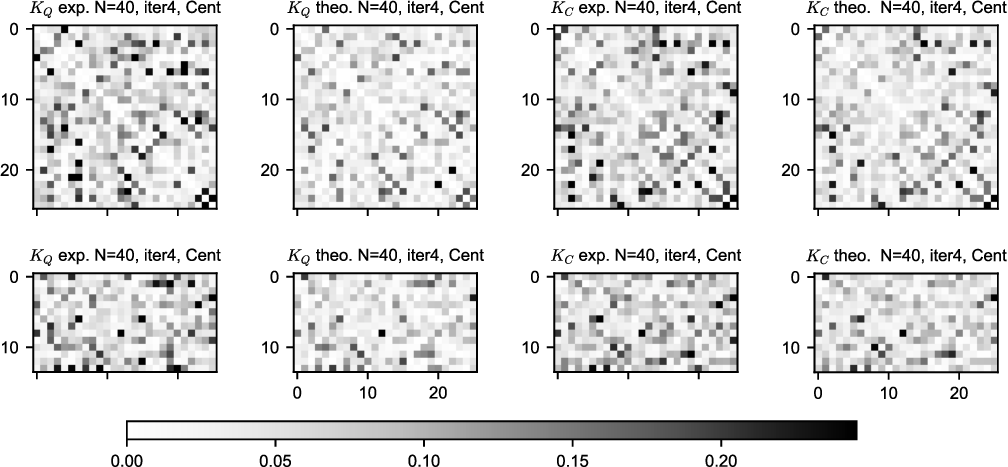}
	\caption{\textbf{Experimentally estimated Gram matrices} We show the Gram matrices corresponding to the quantum kernel (left) and the ones referring to the coherent kernel (right), compared to the theoretical predictions. 
	The number of data points is 40, and the encoding state is $|0,0,1,1,0,0\rangle$. The top row shows the inner products of training gram matrices and the bottom row shows the train-test gram matrices.}
	\label{fig:gram}
\end{figure*}

The probability fidelity registered for the implementation of unitaries are reported in Figure \ref{fig:fid}. The mean fidelity is 0.9816±0.0148 and 0.9934±0.0048, for the quantum kernel and coherent kernel respectively.
To illustrate the difference between two kernels, an example of estimated gram matrices is shown in Figure \ref{fig:gram}.

